\documentstyle[prl,aps,floats,psfig]{revtex}

\def\tp{\hbox{$t_{\perp}$}}

\def\G{{\cal G }}
\def\om{\omega}
\def\iom{i \omega}
\def\kx{{k_{\parallel}}}
\def\ky{{k_{\perp}}}
\def\g{\sigma}

\def\ffour#1{[#1]_4}

\def\v#1{{\bf #1}}

\def\fc#1{\left| #1 \right|}

\def\kom{\kx,\iom;\ky}

\def\ww{{\cal E}}

\def\dnkf{\hbox{$\Delta n_{k_F}\ $}}

\def\dkf{\hbox{$\Delta k_F\ $}}

\def\wbar{\overline}

\def\ch{\rho}

\def\al{\alpha}

\def\kr{K_{\ch}}

\long\def\taglia#1{#1}

\long\def\tagliasi#1{}

\def\beq{\begin{equation}}
\def\eeq{\end{equation}}
\def\beqn{\begin{eqnarray}}
\def\eeqn{\end{eqnarray}}

\def\eqref#1{ %
 (\ref{#1})}

\long\def\singlecol#1{
\twocolumn[\hsize\textwidth\columnwidth\hsize\csname @twocolumnfalse\endcsname
              #1]}

\newcommand{\pcite}[1]{\cite{#1}}

\long\def\beginfigeps#1#2{\begin{figure}[htb] 
   \centerline{\psfig{file=#1,width=8.4cm}}
    \protect{#2}
 \end{figure}}
\long\def\beginfigepstwo#1#2#3{\begin{figure}[htb] %
\par 
\vspace*{-.3cm} 
\par
\centerline{\psfig{file=#1,width=9.4cm}}
\par 
\vspace*{-1.2cm}
\par
\centerline{\psfig{file=#2,width=9.4cm}}
\par 
\vspace*{-.3cm} 
\par
    \protect{#3}
 \end{figure}}
\def\endfigures{}

\ifpreprintsty
      \long\def\beginfigeps#1#2{}
      \long\def\beginfigepstwo#1#2#3{}
      \def\endfigures{
           \long\def\beginfigeps##1##2{\begin{figure}  ##2      \end{figure}}
           \long\def\beginfigepstwo##1##2##3{\begin{figure}  ##3 \end{figure}}
\allfigures}
\fi

\def\diag{\beginfigeps{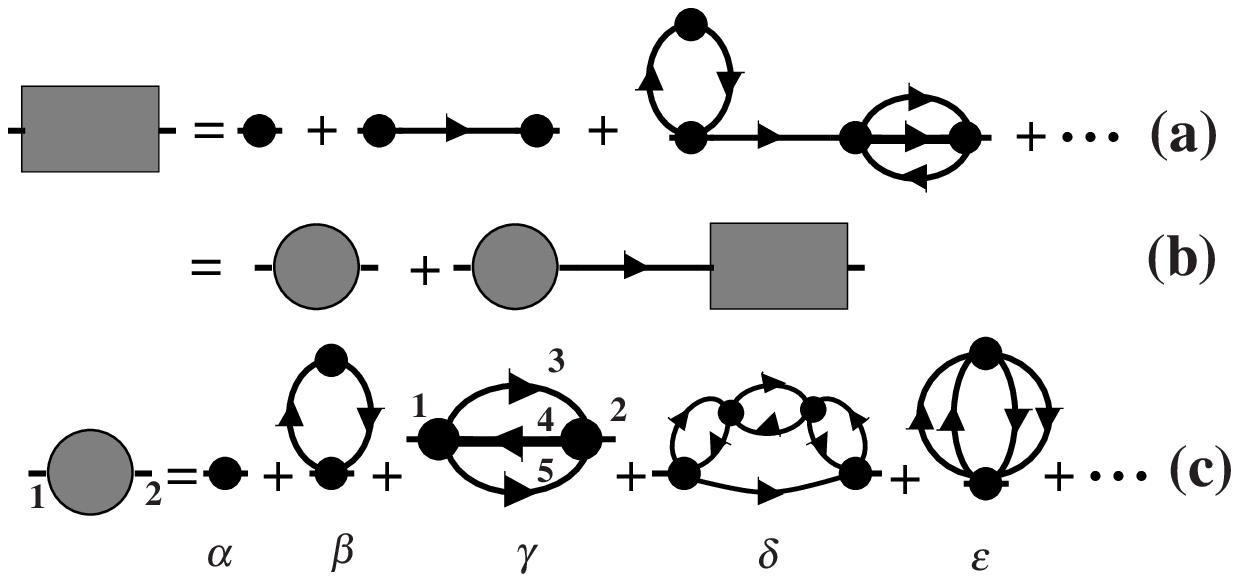}
{\caption{
Diagrammatic expansion in $\tp$ of the single-particle
  Green's function $\G$ (gray box). Directed lines give a contribution
  $\tp[\ky]$, and a dot with $n$ entering and $n$ leaving lines 
  contributes a factor 
$C^0_{n}
$
($n$-particle cumulant).
 (a) Example of single-particle irreducible and
  reducible contributions to $\G$. (b) Dyson's equation for $\G$ in
  terms of the inverse-self-energy $\Gamma$ (gray disk). 
(c) Example of diagrams  contributing to $\Gamma$  (some of them are
 discussed in the text). 
}
\label{diag}}}

\def\expfig{
  \beginfigepstwo{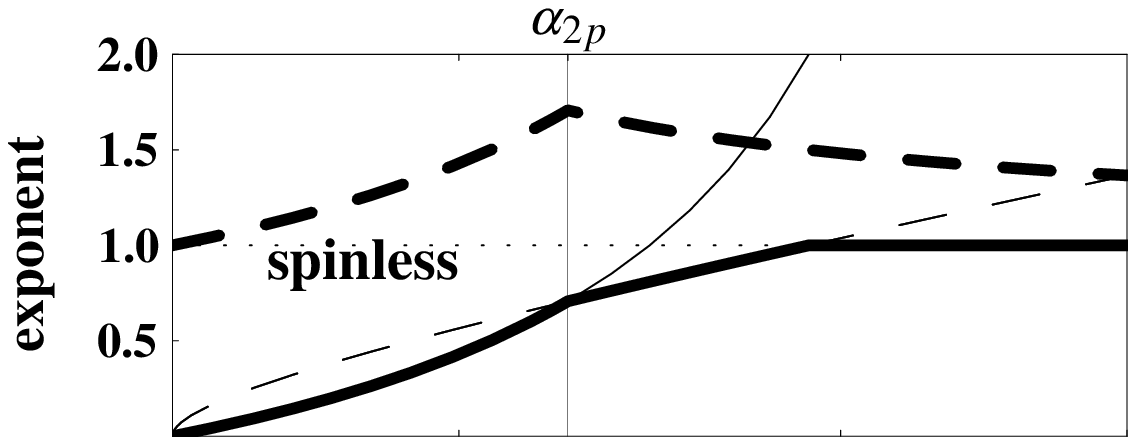}{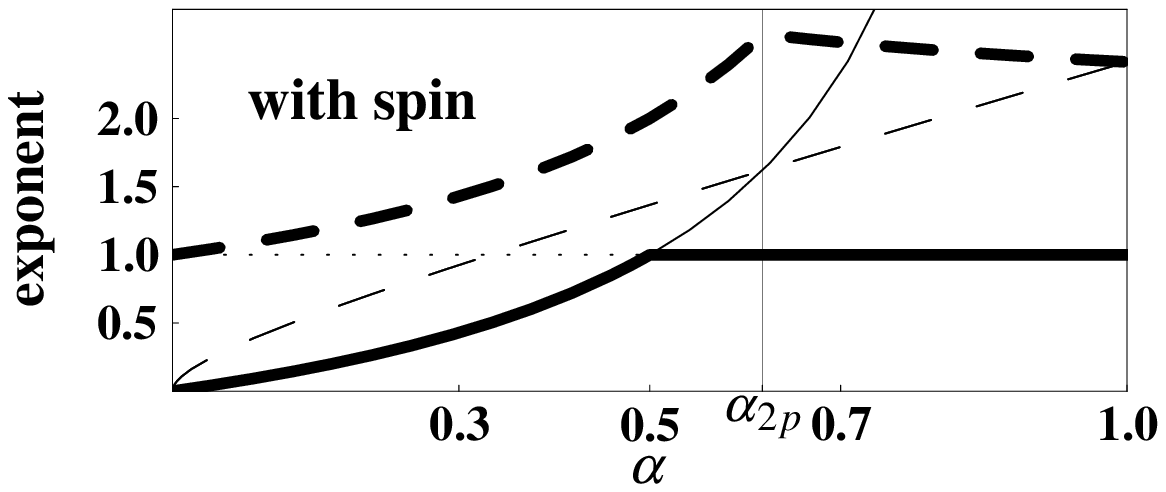}{
\caption{
Exponents controlling the behavior as a function of $\tp$ of 
$\dnkf$ (thick solid line), and 
of the FS 
warp  \dkf (thick dashed line) for the $N$-chains system.
Thin lines indicate  subdominant exponents for $\dnkf$: 
$\al/(1-\al)$ (solid), $\al/(B-\al)$ (dashed), 
and unity (dotted). 
The exponents are plotted as a function of the correlation exponent 
$\al$ for the spinless and
for the spinful cases.
}
\label{expfig}}}

\def\allfigures{
\diag
\expfig
}

\begin{document}

\draft   %

\title{ 
Interchain coherence
of coupled Luttinger liquids at all orders in perturbation
theory 
}

\author{Enrico Arrigoni
}
\address{ 
Institut f\"ur Theoretische Physik,
Universit\"at W\"urzburg,
D-97074 W\"urzburg, Germany
}

\singlecol{

\maketitle

\begin{abstract}
We analyze the problem of Luttinger
liquids coupled via a single-particle hopping $\tp$ and
introduce a systematic diagrammatic expansion 
 in powers of $\tp$.
An analysis of
the scaling of the diagrams 
at each order
allows us to determine
the power-law behavior 
 versus $\tp$
of the interchain hopping and of the Fermi surface warp.
In particular, for strong interactions,
we find that the 
exponents are
 dominated by
higher-order diagrams producing an enhanced coherence and a
failure of  
linear-response theory. 
Our results are
valid at any finite order in $\tp$ for the self-energy.

\end{abstract}

\pacs{PACS numbers : 71.10.Pm,  %
71.10.Hf,  %
05.30.Fk, %
71.15.-m,  %
\hfill{\bf to appear in Phys. Rev. Lett.}
} %
}

The problem of  crossover from one to higher dimensions has
recently
received particular 
interest\pcite{go+cl,wen.90,schu.91,ca.dc.me,bo.bo.95,%
cl.st.96,ho+ko,scho.pr,br.ya.85}.
 The puzzling question
is  whether and how  an infinite row of one-dimensional chains of Luttinger
liquids (LL) 
develops 
 interchain coherence and possibly goes over 
into a Fermi-liquid state for an arbitrarily
weak single-particle hopping
$\tp\ll E_F$ 
 between the chains ($E_F$ being the Fermi energy), and how this is affected 
by the correlation exponent $\al$\cite{conv}  of the unperturbed LL which
parametrizes the {\it intrachain} interaction. 
No clear agreement has been reached so far and different 
analyses have provided conflicting answers:
(i) The system might 
go over to a Fermi-liquid state for
 arbitrarily small $\tp$, or  (ii) there might be a finite $\tp$ (depending on
 $\al$) below which the systems remains in a LL state and no
 coherence is developed between the liquids\pcite{cl.st.96}, 
or finally (iii) one could
 have a different state possibly with gaps and long-range order.
This ambiguity is
essentially due to
the fact that the perturbation introduced by $\tp$ is
 {\it relevant}\pcite{schu.91,ca.dc.me,bo.bo.95}.
This means 
that 
no matter how small $\tp$, perturbative results 
in $\tp$ 
will be always inaccurate for sufficiently low
values of some characteristic
energy $\ww\lesssim\ww_0= (\tp/E_F^{\al})^{1/(1-\al)}$. Here, 
$\ww$ is the 
largest energy scale between (i) the external frequency $\om$, 
(ii) the momentum  measured from the
Fermi surface $|k-k_F|$\pcite{conv}, and (iii) the  temperature $T$.
Indeed,
in the interesting region $\ww\sim\ww_0$, 
 where interchain coherence starts to set on\pcite{bo.bo.95}, 
all perturbation terms
   are of the same order and no definitive prediction can be made
  about the nature of the ground state.
An alternative approach to this topic
is to consider the limit of strong forward scattering
where one can use the higher-dimensional bosonization
method\cite{ho+ko}
or the Ward identities\cite{ca.dc.me}. In this limit, one can show
that 
a system always becomes a
Fermi liquid  in dimensions greater than one if the interaction is not 
too singular.

In this Letter, we 
analyze 
how  {\it interchain coherence} develops
when $\tp$ is switched on and its $\al$ dependence.
The striking 
point of the present work is that the results are valid 
at {\it any finite order} 
 in $\tp$
for the single-particle self-energy, in contrast to previous 
work\pcite{wen.90,bo.bo.95,cl.st.96,scho.pr}.
This is important because it fixes some essentially exact results to
which approximate theories should be compared, although
it is clear that an
answer to the more interesting question ``Fermi liquid or not''
{\it to this degree of accuracy} is extremely difficult and will not be
addressed here.
In addition, we
show  that restricting to  lowest-order terms
may give incorrect results even in the small $\tp$
limit for $\al$  larger than a certain value. In particular,
when considering higher-order diagrams,
 interchain coherence is increased,  and
linear-response theory does not work. 
Specifically, we evaluate the exponent of the power-law behavior 
in $\tp$  (i) of the interchain hopping of a particle at the Fermi momentum,  
 and  (ii) of the Fermi surface (FS) warp.
To calculate these quantities we
extend 
a method developed in Ref. \onlinecite{metz.91} 
to the case of coupled LL
and rewrite
diagrammatically 
the perturbation expansion in $\tp$
introduced in Ref. \onlinecite{bo.bo.95}. 
We then consider the low-energy 
scaling behavior of the diagrams at all orders in
$\tp$ and notice that the scaling is no longer homogeneous
for $\al>\al_{2p}$ (cf. also Refs. \onlinecite{br.ya.85,bo.bo.95}).
Finally, we use this 
scaling behavior to calculate 
the exponents mentioned above.
Recent exact-diagonalization calculations\cite{ed.97} are in excellent agreement
with our predictions.

We consider a system of $N$ coplanar  coupled identical
chains  lying parallel to the $x$-axis with Hamiltonian
$
H = \sum_{i} H_{LL}(i) + 
\sum_{ijr\g} \tp_{ij}  \int d x\ \psi^{\dag}_{r,\g}(x;i) \psi_{r,\g}(x;j) 
$,
where 
$H_{LL}(i)$ describes a
(uncoupled) Luttinger liquid in
chain $i$,
$\tp_{ij}$ is a hopping term between chain $i$ and $j$, and  
$\psi_{r,\g}(x;i)$ ($\psi_{r,\g}^{\dag}(x;i)$) 
is the destruction (creation) operator for a right- ($r=+1$) or
left-moving ($r=-1$) fermion 
at site $x$ in the chain $i$
with spin $\g$ 
(we will also consider the spinless case).
The diagrammatic expansion in $t$
around the atomic
limit of the Hubbard model introduced by W. Metzner 
in Ref. \onlinecite{metz.91}
can be readily extended to the present case by 
considering perturbations in $\tp$ about the exactly-solvable 
Hamiltonian
for the
isolated Luttinger liquids
$\sum_{i} H_{LL}(i)$.
We consider the diagrammatic expansion 
for the Green's function 
$\G(\kom)$ 
expressed in the  
imaginary-frequency ($\iom$)
and 
momentum 
representation, where 
$\kx$ and $\ky$ label the Fourier transforms of 
the $x$ and of the
$i$ components, respectively. 
Unless otherwise specified, $\kx$ will
also implicitly contain the other in-chain quantum numbers  $r$ and $\g$.
The expansion of $\G(\kom)$ in $\tp$  
 is obtained by drawing all connected diagrams composed
of two external lines, and  of an arbitrary number of 
vertices connected with directed lines, 
each vertex having a number of entering lines 
equal to the number of
leaving lines (see Fig. \ref{diag}).
One then labels each internal line $(l)$ by a momentum 
 ($\kx^{(l)}$, $\ky^{(l)}$) 
and a frequency variable ($\iom^{(l)}$), and  the external
lines  by $\kx$, $\ky$ and $\iom$.
 Each {\it internal} line $(l)$ yields a
factor $\tp[\ky^{(l)}]$ (the Fourier transform of $\tp_{ij}$). 
Each vertex with $2n$ legs ($2n\geq 2$) 
 acquires a
contribution $C^0_{n}(\kx^{(1)} \iom^{(1)}, \cdots \kx^{(n)}
\iom^{(n)} |
 \kx^{(1)\prime} \iom^{(1)\prime},
\cdots \kx^{(n)\prime} \iom^{(n)\prime} )$, where $C^0_n$ is the 
$n$-particle cumulant 
of the isolated chain (for example, $C^0_2= \G^0$, the LL Green's
function), 
and $\kx^{(l)} \iom^{(l)}$, $\kx^{(l)\prime}
\iom^{(l)\prime}$
 are the
frequency and momenta (which are conserved at the vertex) 
of the entering and leaving lines, 
respectively  (cf. \pcite{metz.91,bo.bo.95}). 
The introduction of
higher-order cumulants and vertices is necessary, since the
correlation functions of the $\tp=0$ Hamiltonian do not satisfy
Wick's 
theorem.
Eventually, one has to multiply 
each diagram  by 
the usual symmetry and fermion factors
 and integrate (or sum) over all internal frequencies and momenta.
This is a 
diagrammatic representation of the   expansion
introduced by D. Boies and coworkers in Ref. \onlinecite{bo.bo.95}.
Similarly to conventional diagrammatic theory (cf. also
Ref. \onlinecite{metz.91}), one can introduce a 
``self-energy'' $\Gamma(\kom)^{-1}$. The ``inverse-self-energy''
 $\Gamma(\kom)$ is then obtained by summing all
one-particle-irreducible contributions to $\G(\kom)$  (see
Fig. \ref{diag}c).
One then obtains an analogue of the Dyson equation (cf. Fig. \ref{diag}b):
$
\G(\kom) = \left(\Gamma(\kom)^{-1} - \tp[\ky]\right)^{-1} 
$.

At this point,
it is  clear that the approximation used in Refs. 
\onlinecite{wen.90,bo.bo.95,cl.st.96,scho.pr} (and others) consists in
considering only the contribution to $\Gamma$ from the first diagram
(the dot $\al$)  in
Fig. \ref{diag}(c) [``single-dot'' approximation (SDA)] and thus using
$\Gamma=\G^0$.
\taglia{
Furthermore, one can obtain the limit of infinite coordination $D \to \infty$ 
by summing
 an infinite series of ``bow'' diagrams 
with ``full'' self-consistent Green's functions as internal
lines (cf. Ref. \onlinecite{metz.91}).}
\tagliasi{ As pointed out in Ref. 
\onlinecite{bo.bo.95}, the SDA becomes exact if one 
takes the infinite-{\it range} limit
where
  $\tp[\ky] = \tp \delta_{\ky,0}$, which, however, modifies
the Luttinger liquid Green's function 
 only at the special point $\ky=0$.}
Nevertheless,
 we briefly review the results obtained by
 others\cite{wen.90,bo.bo.95,scho.pr}
 within this
SDA.
Here, the system (both spinless and spinful) 
{\it behaves as 
  a Fermi liquid} 
for any values of the correlation exponent $\alpha<1$\pcite{conv},
 in the sense that there is a warped (i. e., $\ky$-dependent) FS $\kx_F[\ky]$,
 on which
 the Green's function has a (real) pole with finite weight at
 $\om=0$ 
and there is a
 {\it finite} region in momentum space 
around $\kx_F[\ky]$ where the pole survives,
 remains real,  and
 continuously shifts to higher binding energies\pcite{fermi}.
 The quasiparticles are thus  well defined
 everywhere on the FS
(except at the {\it special}
  points  for which $\tp[\ky]=0$).
Other poles\cite{cl.st.96} do not describe true quasiparticles, since their
imaginary part is always of the same order as their energy. 
\diag

However, 
the trouble starts to arise when one tries to consider higher-order
diagrams. 
A simple power-counting argument indeed
shows that
a diagram of order $\tp^{n-1}$
for $\Gamma$ diverges  like 
$E_F^{-n \al} \ww^{n(\al-1)}$ in the low-$\ww$ limit. 
If one tries to keep the {\it effective
expansion parameter}  $\tp E_F^{-\al} \ww^{\al-1}$
small, then the $\tp[\ky]$ term in the denominator of $\G(\kom)$ becomes much
smaller than $\Gamma(\kom)^{-1}$ and
the system  behaves as a sum of uncoupled
 LL. 
On the other hand,
 as soon as one tries to approach the pole, 
 the approximation fails and  higher-order diagrams become of the same
 order of magnitude. For this reason, nothing reliable can be said 
about possible Fermi-liquid behavior 
at this order of  perturbation 
 and it is mandatory to consider higher-order diagrams. 
In addition, the above power-counting argument does not hold for all
values of $\al$.
For $\al$ greater than a certain $\al_{2p}$ some diagrams turn out to
diverge stronger at low frequencies 
(cf. also Refs. \onlinecite{br.ya.85,schu.91,bo.bo.95}). For example, the
contribution to 
$\tp\Gamma(\v x_1-\v x_2;\ky)$
\cite{extr}
from the diagram $\gamma$ (Fig. \ref{diag}c), with internal lines $3,4,5$
taking the $r$ indices $+,-,-$, respectively,
 is proportional to 
 \cite{all}
\beqn
\label{scal}
\nonumber
&& \left(\frac{\tp}{E_F^{\al}}\right)^4 \cos \ky  \int 
\prod_{i=3}^5 d^2 \v x_i \  
\left( \fc{\v x_1-\v x_3} \fc{\v x_4-\v x_5} \right)^{-1-\al} 
\\  &&\times
\left[ 
\left(\frac{\fc{\v x_1-\v x_5} \fc{\v x_3-\v x_4}} {\fc{\v x_1-\v x_4} \fc{\v x_3-\v x_5}}\right)^{-B} 
-1 \right] 
\\ \nonumber &&\times
 e^{-i (\arg(\v x_1-\v x_3) - \arg(\v x_4-\v x_5))}
\times
(\hbox{\rm term with } \v x_1 \to \v x_2) \;,
\eeqn 
where $\fc{\v x} \equiv \sqrt{x^2+(|\tau|+a)^2}$, $\arg{\v x}\equiv
 \arg(x + i \tau)$, and 
the exponent $B= (1/\kr-\kr)/(2S)$.
A naive dimensional analysis  
 yields  a term  $ \propto 
\left({\tp}/{E_F^{\al}}\right)^4 
\fc{\v x_1-\v x_2}^{2-4 \al}$,
giving a contribution 
proportional to 
$\left({\tp}/({E_F^{\al} \ww^{1-\al}})\right)^4 $
to the Fourier transform $\tp \Gamma(\kom)$. 
This is  correct for $\ww \ll E_F$ 
if the integral does not diverge at small
distances for $a\to0$,
i.e. as long as $B<1$. For $B>1$\cite{conv} 
one picks up an $a$-dependent
contribution $a^{2-2B}\sim E_F^{2B-2} $, 
which must be balanced by  $\ww^{2-2B}$ to give
the correct energy dimensions. This produces a contribution to  
$\tp \Gamma(\kom)$ proportional to 
$\left({\tp}/({E_F^{\al}
    \ww^{1-\al}})\right)^4
          (E_F/\ww)^{2B-2} $\taglia{, i.e. a stronger divergence}. 
A similar analysis shows that the
$7$-legs diagram  $\delta$ in Fig. \ref{diag}(c)  and its generalizations 
with $2n +1$ internal legs ($n$ odd integer) 
 produce a contribution to $\tp \Gamma$ proportional to
$\left({\tp}/({E_F^{\al} \ww^{1-\al}})\right)^{2n+2}
          (E_F/\ww)^{2n(B-1)} $. 
\taglia{ In fact,
this is the
strongest low-energy divergence one can attain for a contribution 
to $\tp \Gamma$
at a given order $\tp^{2n+2}$.}
The regime $B>1$ corresponds to $\al>\al_{2p}$ with
$\al_{2p}=(\sqrt{S^2+1}-1)/S$,
i.e.,  where  
    two-particle processes become more relevant than 
   one-particle processes\pcite{bo.bo.95,br.ya.85}.

We now use the above results to
calculate the power-law behavior for small $\tp$ of the quantity\cite{extr} 
\beq
\label{dn}
 \dnkf 
\propto  \int_{-\infty}^{\infty} d \om 
\left[\G(k_F,\iom;\pi)-\G(k_F,\iom;0) \right] \;.
\eeq
$ \dnkf $ is
also equal to the expectation value of the single-particle hopping
operator
at $k_F$ in the two-chains case and is thus
a measure of the {\it coherence}\cite{cl.st.96}
 of single-particle hopping.
For $k=k_F$ and $T=0$ the only remaining energy scale that can be
associated with $\ww$ is $\om$.
For $\al<\al_{2p}$ all diagrams scale in the same way as discussed above.
Therefore, {\it at any order in the $\tp$ expansion} and for  $\om \ll E_F$
 the difference between the Green's function 
of the two bands 
$\G(k_F,\iom;\pi) - \G(k_F,\iom;0)$ 
can be written 
(say, for $\hbox{Im} \ \iom>0$) 
 as 
$E_F^{-\al} (\iom)^{\al-1} g\left[ \tp  E_F^{-\al} (\iom)^{\al-1} \right]$,
 where $g[x]$ is a {\it scaling} function.
The    change of variables 
$\om = x \ \tp^{1/(1-\al)}$
gives
$ \dnkf \propto (\tp/E_F)^{\al/(1-\al)} $
times a dimensionless integral, which converges at large $x$
for $\al<1/2$.
To assure convergence at low frequencies it is essential to expand the
{\it self-energy} ($\Gamma^{-1}$) in \tp and not directly the Green's function,
in order to avoid
an {\it unphysical} 
 low-frequency (or low-$T$) divergence.
 For $\al>1/2$ the integral diverges
 for large $x$ and one needs to introduce 
 a large-energy
 cutoff yielding a dominant linear behavior
 $\dnkf \propto \tp/E_F$.
The same exponents would 
 have been obtained by simply using
the SDA for $\Gamma$.
\taglia{Notice also that this result is consistent with the one in Ref. 
\onlinecite{ca.dc.me}.}
 
In the  $\al>\al_{2p}$ regime, 
one cannot straightforwardly extend the above
discussion 
since the diagrams give nonhomogeneous contributions.
If one stops the expansion in $\tp$ for the {\it self-energy}
 at an arbitrary finite order
$\tp^{m_0+1}$
one obtains
(restricting to dominant terms at each order)
\beqn 
\label{gsc}
&&
\G(k_F,\iom;\pi)^{-1} 
\\ \nonumber &&
= 
\tp \sum_{m=-1}^{m_0} a_m \left( (\iom)^{\al-1} E_F^{-\al} \tp \right)^m
        \left(\frac{\iom}{E_F}\right)^{(1-B) \ffour{m}}\;,
\eeqn
where $\ffour{m}$ is defined as
$\ 4\ {\rm int}((m-2)/4)+2$, int being the (lower) integer part, 
and the  $a_{m}$
are constants ($a_0=1$) possibly dependent on the spin
and charge velocities.
$\G(k_F,\iom;0)^{-1}$ is obtained from the same expression by replacing
$\tp \to
 -\tp$.
By inserting \eqref{gsc} into the integral \eqref{dn} 
 one can show that 
 the integral is dominated
by the constant 
 and by the 
$\tp^{\wbar m_0+1}$ terms in \eqref{gsc}.
 Carrying out the integral yields
$\dnkf \propto \tp^{R[\wbar m_0]}$, where the exponent
$R[\wbar m_0]$
 approaches 
 quite rapidly its $\wbar m_0 \to \infty$ limit $R= \al/(B-\al)$.
For $R>1$ this term will again be shaded by the linear term as in the
case $\al<\al_{2p}$.
 \expfig
In Fig. \ref{expfig}, we 
 show the {\it dominant} and
some subdominant exponents for the spinless and spinful cases as a function
of $\al$. 
As one can see, the occurrence of the two-particle 
regime ($\al>\al_{2p}$)
reduces \
the dominant exponent in the spinless case, 
and, in particular, it shifts upwards
(from $\al >1/2$ to $\al>2/3$)
the region where $\dnkf$ becomes linear in $\tp$.
It is interesting to compare this result with the 
prediction of linear-response theory, which 
is expected, for
sufficiently small $\tp$, either to give the correct behavior or to
diverge thus
signaling a sublinear behavior.
The linear-response result 
$\Delta n_k = \tp ( A + B |\kx-k_F|^{2 \al -1})$ 
indeed shows a
divergence at $\kx = k_F$ for $\al<1/2$, i. e. it predicts a sublinear
behavior in this region.
However, our calculation shows 
  that  in the spinless case $\dnkf$ is sublinear in $\tp$ in a {\it larger}
  region, namely, up to $\al=2/3$.
Unfortunately, this effect is more difficult to
see in the  case of electrons 
with spin, since in this case $\al_{2p} \approx 0.62 > 1/2$
and  the effect is ``shaded'' by the linear
behavior occurring for $\al>1/2$.
\taglia{Nevertheless,  this effect should be detectable, e.g., in the
first derivative of $\dnkf$
with respect to $k$ or in the FS warp 
as discussed below.}
Notice that the exponent $R$
has been obtained by cutting the series 
 in $\tp$ {\it for the self-energy $\Gamma^{-1}$} at a given order.
For $\al>\al_{2p}$,
a different choice, like, e.g., cutting the series {\it for $\Gamma$},
 may lead to a different (although  in this case unphysical) result,
which, however, preserves
 the qualitative effects, in particular the decrease of  the exponent $R$.
It should be mentioned that 
recently S. Capponi, D. Poilblanc and myself\pcite{ed.97}, 
have evaluated 
the exponent of \dnkf  by exact diagonalization of  small
ladders supplemented by a careful finite-size extrapolation
carried out by means of an appropriate scaling function.
These numerical results are in very good accordance with the exponents 
for $\dnkf$ predicted here
and clearly show the change of behavior between the two
regimes  $\al<\al_{2p}$ and  $\al>\al_{2p}$. 

 We  now study the 
behavior of the
Fermi surface as a function of $\tp$. 
The FS
consists of the points $\kx_F[\ky]$ given by the solution of the 
equation\cite{conv}
$\hbox{Re}\ \G(\kx_F[\ky],\iom=0^+;\ky)^{-1}=0$. As already 
discussed\pcite{wen.90,bo.bo.95,scho.pr}, within the
SDA a solution exists 
and gives
$\kx_F[\ky]-k_F \propto \left|\tp[\ky]\right|^{1/(1-\al)}$. 
At higher order,
$\G(\kx,\iom=0^+;\pi)^{-1}$ has the same form as \eqref{gsc} with $\iom$
replaced with $\kx-k_F$ and with different coefficients $a_n$, whereby
$B$ is formally replaced  with $1$
for $\al<\al_{2p}$.
The equation for the ($\ky=\pi$) 
pole can be written in terms of a scaling function $f$ as
$f[\tp/(\kx-k_F)^{1-\al}]=0$.
If $f[x]=0$ has a solution at some point $x=x^*$, then the behavior of
the FS
is similar to the one for the SDA,
i.e. $\dkf \equiv \kx_F[\pi]-\kx_F[0]
= (\tp/x^*)^{1/(1-\al)}$ \pcite{kpos}. 
This quantity measures
the FS warp 
produced by $\tp$.
However, it is not easy to find out 
whether the imaginary part of
$\G^{-1}$ vanishes fast enough at the 
FS, so that the above
result does not necessarily imply that the system goes into  a Fermi
liquid state.
Moreover,  considering higher-order terms
for $f[x]$, the solution of 
$f[x]=0$
 may not exist. 
In fact, several gaps open in the different modes of a $N$-chain
system\pcite{libafi}, which 
could possibly prevent the pole equation from having a solution
for some $N$ and some $\ky$ points of the FS.
In this case, as a measure of the FS warp,
 one could
define a ``pseudo'' FS by the point
$\kx$ 
for which 
$|\G^{-1}|$ has a minimum.  
Also in this case, the FS warp turns out to
behave like  $\tp^{1/(1-\al)}$.
Finally, for $\al>\al_{2p}$ the behavior for small $\tp$ is
again dominated by the
dot diagram ($\alpha$) and by the higher-order diagrams of type
$\gamma$ and $\delta$, as
for the calculation of $\dnkf$. In a similar way, one 
obtains for the 
FS warp
$\dkf \propto \tp^{R/\al}$ (cf. Fig. \ref{expfig}).
This means that the Fermi surface is warped
even for
$\al \to 1$ in contrast with the SDA result and with the 
expectation  coming form the fact that $\tp$
is ``irrelevant'' for $\al>1$.
Notice, however, that
particle-hole instabilities 
producing a gap at the Fermi surface
are likely to occur for large $\al$.

In conclusion, using a 
 diagrammatic representation of the 
expansion in
power of $\tp$, 
we have determined the power-law 
behavior   as a function of $\tp$
of the difference in
occupation  $\dnkf$ 
and of the FS warp
in a $N$-chain system, 
our results being exact at {\it any finite order} for the self-energy.
 In the single-particle regime
($\al<\al_{2p}$),
the exponents are correctly given by the low-order  approximations. In the two-particle
regime ($\al>\al_{2p}$) higher-order diagrams 
give dominant contributions and reduce
 the exponents (thus indicating an increased coherence) with respect to the
SDA result. In this regime, linear-response
theory in $\tp$ is not reliable
 {\it even in the small-$\tp$} limit.
The increase of the exponents with $\al$, as seen in Fig.  \ref{expfig},
shows that transverse coherence is reduced with increasing
 interaction but never completely suppressed since
\dnkf and the Fermi surface warp are
 finite for any finite value of $\tp$ also for large $\al$.
We expect, however, that at some value of $\al$, 
a gap opens due to particle-hole instabilities possibly producing
 incoherence effects in the intrachain transport.
For a finite number of chains these gaps seems to open  even for small
 $\al$\pcite{libafi}. 
It would be interesting to establish 
whether a
finite  critical value of $\al$ exists in the $N\to \infty$ limit.

\nocite{voit.rev}
I'm grateful to J. Voit for 
stimulating my interest in 
the topic of
quasi-one-dimensional systems. I also acknowledge many interesting and
fruitful discussions with M. G. Zacher, S. Capponi, D. Poilblanc, B. Brendel,
G. Hildebrand
 and W. Hanke.
This work was supported by the 
EC-TMR
  program  ERBFMBICT950048.

\long\def\bibdef#1#2#3{\newenvironment{#1}{#2}{}}
\def\lcit#1{\begin{#1}\end{#1}}
\ifx\undefined\andword \def\andword{and} \fi %
\ifx\undefined\submitted \def\submitted{submitted} \fi %
\ifx\undefined\inpress \def\inpress{in press} \fi %
\def\nonformale#1{#1}
\def\formale#1{}
\def\spa{} \def\spb{}
\def\spa{\ifpreprintsty\else\vspace*{-.5cm}\fi} %
\def\spb{\ifpreprintsty\else\vspace*{-1.6cm}\fi} %

\bibdef{schu.96}{
H.~J. Schulz, Phys. Rev. B {\bf 53},  R2959  (1996)}%

\bibdef{ho.ma.93}{
A. Houghton and J.~B. Marston, Phys. Rev. B {\bf 48},  7790  (1993)}%

\bibdef{ko.me.93}{
P. Kopietz, V. Meden, and K. Sch{\"o}nhammer, Phys. Rev. Lett. {\bf 74},  2997
  (1995)}%

\bibdef{cl.st.94}{
D.~G. Clarke, S.~P. Strong, and P.~W. Anderson, Phys. Rev. Lett. {\bf 72},
  3218  (1994)}%

\bibdef{go.dz.74}{
L.~P. Gor'kov and I.~E. Dzyaloshinskii, Sov. Phys. JETP {\bf 40},  198
  (1974)}%

\bibdef{li.ba.97}{
H.-H. Lin, L. Balents, and M.~P. Fisher, Phys. Rev. B {\bf \it in press},
  (1997), (cond-mat/9703055)}%

\def\nonformale#1{#1}
\def\formale#1{}
\spa

\endfigures


\begin{thebibliography}{10}
\bibitem{extr}{
For definiteness, we consider the right-moving branch with external index
  $r=+$.}%
\bibitem{all}{
Here, $\Gamma$ is expressed in real space and imaginary time for the interchain
  coordinates $\v x\equiv (x,i \tau)$ and in momentum space for the intrachain
  coordinate. For simplicity, we will set the spin and charge velocities equal
  to $v_F$ $(=1)$. This does not change the scaling behavior we are
  considering. In the following, the index $S$ equals $1$ ($2$) in the spinless
  (spinful) case, $a\propto E_F^{-1}$ is a short-distance cutoff, and $T=0$.
  Moreover, we take $N$ even and restrict $t_{ij}$ to nearest neighbors. Notice
  that the $-1$ in square brackets in Eq. \eqref{scal} comes from the
  disconnected part that has to be subtracted in order to obtain the cumulant
  $C^0_4$. This term is important otherwise the integral would diverge in the
  infinite-volume limit.}%
\end{thebibliography}
\end{document}